\documentclass{aa}
\usepackage{graphicx}
\usepackage{natbib}
\bibpunct{(}{)}{;}{a}{}{,}
\usepackage{wasysym}

\usepackage{amsmath}
\usepackage{cancel}
\usepackage{multirow}
\usepackage{float}
\usepackage{array} 
\usepackage{tikz}
\usepackage{xcolor}
\usepackage{color}
\usepackage{physics}
\usepackage{algpseudocode}
\usepackage{url}
\usepackage{amsfonts,amssymb}
\usepackage{soul}
\usepackage[switch]{lineno}

\newcommand{\ve}[1]{{\rm\bf {#1}}}

\begin{document}

\title{Inference of horizontal velocity fields from the induction equation in the solar atmosphere}
\subtitle{I. Analytical and numerical solutions in 2D}
\author{H.~Vila Crespo\inst{1,2} \and J.M.~Borrero\inst{1} \and I.~Mili\'c\inst{1} \and {G.~Vigeesh}\inst{1} \and A.~Asensio Ramos\inst{2,3}}
\institute{Institut f\"ur Sonnenphysik, Georges-K\"ohler-Allee 401a, D-79110, Freiburg, Germany
\and
Departamento de Astrof{\'\i}sica, Universidad de La Laguna, E-38205, La Laguna, Tenerife, Spain 
\and
Instituto de Astrof{\'\i}sica de Canarias, Avd. V{\'\i}a L\'actea s/n, E-38205, La Laguna, Tenerife, Spain
}
\date{Received 12 December 2025 / Accepted 17 February 2026}

\abstract{Spectroscopic and spectropolarimetric observations, which rely on the Doppler effect, only provide access to the line-of-sight component of the solar plasma velocity (i.e., $v_z$). However, many dynamic processes in the solar atmosphere involve strong horizontal motions (i.e., in the plane perpendicular to the line of sight: $v_x$, $v_y$). Existing methods for estimating horizontal velocities are generally insensitive to variations in height (i.e., the $z$-coordinate), providing them only on a single plane perpendicular to the line of sight: $v_x(x,y)$, $v_y(x,y)$.}{Motivated by the fact that modern analysis techniques (i.e., Stokes inversion) allow us to retrieve the height dependence of $v_{z}$ and ${\bf B}$, our goal is to infer also this height dependence for the horizontal velocity field in the solar atmosphere. As a first step, we present, develop, and test a method for the two-dimensional case on the $(y,z)$ plane so as to show that the $z$ dependence can be successfully retrieved.}{The components of the two-dimensional magnetic induction equation are discretized via finite differences, leading to an overdetermined system whose solution provides $v_y(y,z)$. The method assumes that $\bf{B}$, its time variation $\dot{\bf{B}}$, as well as $v_z$ are known. This is currently possible through modern Stokes inversion techniques applied to spatially and temporally resolved spectropolarimetric observations.}{Using analytically prescribed values and two-dimensional magnetohydrodynamic simulations of the solar surface, we demonstrate that, in these idealized cases, the horizontal velocity component in a two-dimensional domain, $v_y(y,z)$, can be successfully recovered with a mean error of about 1~\%. We observe that in the regions where either the modulus of the velocity or its horizontal components are close to zero, its retrieval worsens in comparison to the rest of the domain.}{The proposed method successfully retrieves the horizontal velocity field in the $(y,z)$ plane, thereby establishing the foundation for future extensions to three-dimensional reconstructions of the horizontal velocity field.}

\titlerunning{Induction-based inference of horizontal velocities in the solar atmosphere: 2D cases}
\authorrunning{Vila Crespo et al.}
\keywords{Sun: magnetic fields -- Sun: photosphere -- Magnetohydrodynamics (MHD)}
\maketitle

\section{Introduction}
\label{section:introduction}

\noindent Through the widely employed techniques of spectroscopy and spectropolarimetry (which rely on the Doppler effect), it is only possible to infer the line-of-sight velocity of solar plasma. For simplicity, we refer to this velocity component, using Cartesian coordinates, as $v_z$ at the disk center. However, many interesting phenomena that occur in the solar atmosphere involve the presence of large velocities in the plane perpendicular to the observer's line of sight. We refer to these as $v_x$ and $v_y$ components of the velocity field. Some of these phenomena include flows at a sub-granular scale \citep{steiner2010granulation}, the presence of swirling motions \citep{bonet2010lct_vortex}, diverging flows from magnetic reconnection events \citep{thaler2023photosphericreconnection}, and magnetic tornadoes \citep{wedemeyer2012magnetictornadoes}. Apart from these phenomena, determining horizontal velocities ($v_{x}, v_{y}$) is extremely important because they can be employed to estimate the flux of magnetic energy and helicity from the photosphere into the corona \citep{welsch2007comparison, welsch2014active, kazachenko2015fluxes, tilipman2023quiet} as well as to improve the inference of electric currents in the solar atmosphere \citep{adur2021electriccurrents,borrero2023electriccurrents}. 

The most widely used technique to determine velocities in the solar atmosphere in the plane perpendicular to the line of sight is the local correlation tracking technique \citep[LCT;][]{november1988lct,simon1988lct}. This is an optical technique that uses time-resolved monochromatic intensity images, typically at a continuum wavelength where no spectral lines are present. The LCT has been successfully employed to determine motions on the solar surface in a variety of photospheric structures, such as granulation \citep{simon1995lct_gra}, pores \citep{johann2002lct_pore}, sunspot penumbra \citep{denker1998lct_pen,sobotka2001lct_pen}, sunspot moats \citep{vargas2008lct_pen}, and vortices \citep{bonet2010lct_vortex}, among others. Newer techniques include the use of deep learning neural networks \citep{andres2017lct_nn} and a local correlation tracking technique based on Fourier analysis (FLCT), which has been applied to time-resolved images of the continuum intensity \citep{fisher2008flct} as well as to time-resolved maps of the line-of-sight component of the magnetic field $B_z$ \citep{welsch2004flct, deforest2007tracking, welsch_2015_hinodepoyntingflux,kostic2025}.

Assuming that the horizontal velocity inferred using LCT, referred to as $\ve{u}_{h}$, is equivalent to real plasma velocities $\ve{u}_{h} = \ve{v}_{h}$ \citep{kusano2002induction} or related to them via $\ve{u}_{h} = \ve{v}_h - (v_z/B_z) \ve{B}_{h}$, for example, as proposed by \cite{demoulin2003uf}, the vertical component (i.e., $z$-component) of the induction equation in ideal magnetohydrodynamics (MHD) can be incorporated to determine the full velocity vector $\ve{v}$ using different techniques \citep{welsch2004flct,longcope2004induction,schuck2008magnetograms}. These are referred to as inductive techniques, and they assume that the time derivative of $B_z$ is known (i.e., obtained from a time series of line-of-sight magnetograms).

All the aforementioned methods share a common feature: They all retrieve the full velocity vector\footnote{Throughout this paper, vector quantities are indicated in bold roman letters, whereas components along each spatial axis are in italics: $\ve{A}=(A_x,A_y,A_z)$. Matrices are represented by the symbol $\widehat{\bf A}$, and their elements are referred to as $\widehat{A}_{ij}$} $\ve{v}$ in a $(x,y)$ plane at a fixed height (or single $z$ height) on the solar atmosphere: $\ve{v}(x,y)$. In this paper, we present an extension and modification of these methods where all three components of the induction equation are employed, enabling the inference of the horizontal components of the velocity at multiple heights in the solar atmosphere: $\ve{v}_{h}(x,y,z)$.

This new method presumes that the full magnetic field vector $\ve{B}(x,y,z,t)$ and the line-of-sight component of the plasma velocity $v_z(x,y,z,t)$ are known. Nowadays, these physical parameters can be inferred, both in the photosphere \citep{borrero2025firtez} and chromosphere \citep{anjali2023ellerman}, through advanced inversion techniques for the polarized radiative transfer equation \citep{adur2019firtez,borrero2021firtez} when applied to observations of the Stokes vector in photospheric and chromospheric spectral lines as a function of time, $\ve{I}(x,y,\lambda,t)$, and carried out with modern spectropolarimeters \citep{dominguez2022ifu,vannoort2022mihi}. 

In this paper, we present the first application of this newly developed method to the two-dimensional velocity and magnetic vector fields, both confined to the $(y,z)$ plane. The description of the method is provided in Sect.~\ref{section:numerical_method}. In Sect.~\ref{subsection:no_temporal_yz} we apply it to analytically prescribed solutions for the velocity and magnetic fields, whereas in Sect.~\ref{subsection:simulation_yz} we apply it to a two-dimensional MHD simulation of the solar surface. A discussion on the errors in the inference of horizontal velocities in two dimensions is presented in Sect.~\ref{section:discussion}. Finally, we present a summary of our findings in Sect.~\ref {section:conclusions}.

\section{Numerical method}
\label{section:numerical_method}

\subsection{General remarks}
\label{subsection:general_remarks}

Our ultimate goal is to infer the horizontal velocities $v_{x}, v_{y}$ (i.e., those perpendicular to the line of sight direction). To this end, we used the induction equation under the assumption of ideal MHD (i.e., zero magnetic diffusion):

\begin{eqnarray}
\frac{\partial \ve{B}}{\partial t} = \nabla \times (\ve{v} \times \ve{B}) \;.
\label{eq:induction_eq_vect}
\end{eqnarray}

This equation describes how the magnetic field evolves over time in the presence of a velocity field and is satisfied at all times. Therefore, we can specify a particular instant in time, $t_0$:

\begin{eqnarray}
\eval{\frac{\partial \ve{B}}{\partial t}}_{t_0} = \dot{\ve{B}}(t_0) = \nabla \times \left (\ve{v}(t_0) \times \ve{B}(t_0) \right) \;.
\label{eq:induction_eq_vect_eval}
\end{eqnarray}

As mentioned in Sect.~\ref{section:introduction}, it is possible to observationally infer ${\bf B}, \dot{\ve{B}}$ and $v_{z}$. Hence the solution of this equation will provide the horizontal components (i.e., perpendicular to the line of sight) of the velocity: $v_{x}$ and $v_{y}$.

\subsection{Two-dimensional induction equation in the $(y,z)$ plane}
\label{subsection:induction_equation_yz}

\noindent Although the remarks in Sect.~\ref{subsection:general_remarks} apply to the general three-dimensional case, the goal of this paper is to first test our method in two dimensions, the $(y,z)$ plane, where the $z$-direction corresponds to the observer's line of sight. To this end, we set $v_x = B_x =0$, and thus the horizontal velocity to be retrieved becomes $v_y(y,z)$. Under these assumptions, the two nontrivial components of the induction equation can be written as

\begin{eqnarray}
\dot{B_y} & = & \frac{\partial}{\partial z} (v_{y} B_{z} - v_{z} B_{y}) 
\label{eq:induction_eq_2D_yz_y} \\
\dot{B_z} & = & \frac{\partial}{\partial y} (v_{z} B_{y} - v_{y} B_{z})
\label{eq:induction_eq_2D_yz_z},
\end{eqnarray}

\noindent where for simplicity $t_0$ is not indicated anymore. We further assumed that the Stokes inversion provides $B_y$, $B_z$, $\dot{B_y}$, $\dot{B_z}$, and $v_z$. With this, we obtained an overdetermined system with one unknown, $v_y$, and two equations for each point of the two-dimensional domain.

\subsection{Discretization and matrix form}
\label{subsection:discretization_matrix}

\noindent We used the finite differences method with a five-point centered stencil throughout the entire domain to solve the induction equation numerically. With this, the first derivative of a continuous function, $f$, along direction $r$ at point $i$ can be written as

\begin{eqnarray}
\frac{\partial f}{\partial r} \bigg|_{i} = \frac{f_{i-2} - 8f_{i-1} + 8f_{i+1} - f_{i+2}}{12h_r} 
\label{eq:first_derivative_5_point},
\end{eqnarray}

\noindent where $h_r$ is the stencil length along direction $r$. Applying this stencil to Eqs.~\ref{eq:induction_eq_2D_yz_y} and~\ref{eq:induction_eq_2D_yz_z}, we obtained

\begin{eqnarray}
\begin{split}
\dot{B_y}(i,j) + & \eval{\frac{\partial}{\partial z}(v_{z}B_{y})}_{i,j} = \frac{B_{z}(i,j)}{12 h_{z}} v_{y}(i,j-2) \\ &  - \frac{2 B_{z}(i,j)}{3 h_{z}} v_{y}(i,j-1) + \eval{\frac{\partial B_{z}}{\partial z}}_{i,j} v_{y}(i,j) \\ & + \frac{2 B_{z}(i,j)}{3 h_{z}} v_{y}(i,j+1) -\frac{B_{z}(i,j)}{12 h_{z}} v_{y}(i,j+2)
\end{split}
\label{eq:induction_eq_2D_yz_y_5p}
\end{eqnarray}
\begin{eqnarray}
\begin{split}
\dot{B_z}(i,j)  - & \eval{\frac{\partial}{\partial y}(v_{z}B_{y})}_{i,j} =  - \frac{B_{z}(i,j)}{12 h_{y}}v_{y}(i-2,j) \\ & + \frac{2 B_{z}(i,j)}{3 h_{y}} v_{y}(i-1,j) - \eval{\frac{\partial B_{z}}{\partial y}}_{i,j} v_{y}(i,j) \\ & - \frac{2 B_{z}(i,j)}{3 h_{y}} v_{y}(i+1,j) + \frac{B_{z}(i,j)}{12 h_{y}} v_{y}(i+2,j),
\end{split}
\label{eq:induction_eq_2D_yz_z_5p}
\end{eqnarray}

 \noindent where $\eval{\frac{\partial B_{z}}{\partial z}}_{i,j}$ and $\eval{\frac{\partial B_{z}}{\partial y}}_{i,j}$ are also evaluated according to Eq.~\ref{eq:first_derivative_5_point}. Equations~\ref{eq:induction_eq_2D_yz_y_5p} and ~\ref{eq:induction_eq_2D_yz_z_5p} provide the basis to write an overdetermined linear system of equations where $v_y(i,j)$ are the unknown quantities. The coefficients that multiply the unknown $v_y$ in the equations above are renamed $a_{i,j}$ or $b_{i,j}$, depending on whether they appear in the equations pertaining to $\dot{B_y}$ or $\dot{B_z}$, respectively. These coefficients are also given a superscript that refers to the grid point at which the horizontal component of velocity they multiply is evaluated. In addition, the independent coefficients (i.e., those that do not multiply $v_y$) are renamed $c^{a}_{i,j}$ and $c^{b}_{i,j}$. The superscript $a$ or $b$ indicates whether the independent coefficient appears in the equation for $\dot{B_y}$ or $\dot{B_z}$,. From Equation~\ref{eq:induction_eq_2D_yz_y_5p}, we defined

\begin{eqnarray}\label{eq:ac_coeffs}
a_{i,j}^{i,j-2} & = & \frac{B_z(i,j)}{12 h_z} \notag\\
a_{i,j}^{i,j-1} & = & - \frac{2 B_z(i,j)}{3 h_z} \notag\\
a_{i,j}^{i,j} & = & \eval{\frac{\partial B_{z}}{\partial z}}_{i,j}\\
a_{i,j}^{i,j+1} & = & \frac{2 B_z(i,j)}{3 h_z}\notag\\
a_{i,j}^{i,j+2} & = & - \frac{B_z(i,j)}{12 h_z} \notag\\
c_{i,j}^{a} & = & \dot{B_y}(i,j) + \eval{\frac{\partial}{\partial z}(v_{z}B_{y})}_{i,j}\notag.
\end{eqnarray}

Likewise, for the $b_{i,j}$ and $c_{i,j}^b$ coefficients appearing in Equation~\ref{eq:induction_eq_2D_yz_z_5p},

\begin{eqnarray}\label{eq:bc_coeffs}
b_{i,j}^{i-2,j} & = & -\frac{B_z(i,j)}{12 h_y} \notag\\
b_{i,j}^{i-1,j} & = & \frac{2 B_z(i,j)}{3 h_y} \notag\\
b_{i,j}^{i,j} & = & -\eval{\frac{\partial B_{z}}{\partial y}}_{i,j}\\
b_{i,j}^{i+1,j} & = & -\frac{2 B_z(i,j)}{3 h_y}\notag\\
b_{i,j}^{i+2,j} & = & \frac{B_z(i,j)}{12 h_y} \notag\\
c_{i,j}^{b} & = & \dot{B_z}(i,j) - \eval{\frac{\partial}{\partial y}(v_{z}B_{y})}_{i,j}\notag.
\end{eqnarray}

With this, Eqs.~\ref{eq:induction_eq_2D_yz_y_5p} and ~\ref{eq:induction_eq_2D_yz_z_5p} become

\begin{eqnarray}\label{eq:induction_eq_2D_yz_y_5p_simpl}
\begin{split}
c^{a}_{i,j} = & a_{i,j}^{i,j-2} \cdot v_{y}(i,j-2) + a_{i,j}^{i,j-1} \cdot v_{y}(i,j-1) \\ & + a_{i,j}^{i,j} \cdot v_{y}(i,j) + a_{i,j}^{i,j+1} \cdot v_{y}(i,j+1) \\ & + a_{i,j}^{i,j+2} \cdot v_{y}(i,j+2)
\end{split}
\end{eqnarray}

\begin{eqnarray}\label{eq:induction_eq_2D_yz_z_5p_simpl}
\begin{split}
c^{b}_{i,j} = & b_{i,j}^{i-2,j} \cdot v_{y}(i-2,j) + b_{i,j}^{i-1,j} \cdot v_{y}(i-1,j) \\ & + b_{i,j}^{i,j} \cdot v_{y}(i,j) + b_{i,j}^{i+1,j} \cdot v_{y}(i+1,j) \\ & + b_{i,j}^{i+2,j} \cdot v_{y}(i+2,j).
\end{split}
\end{eqnarray}

\begin{figure}[h]
    \centering
    \includegraphics[width=0.7\linewidth]{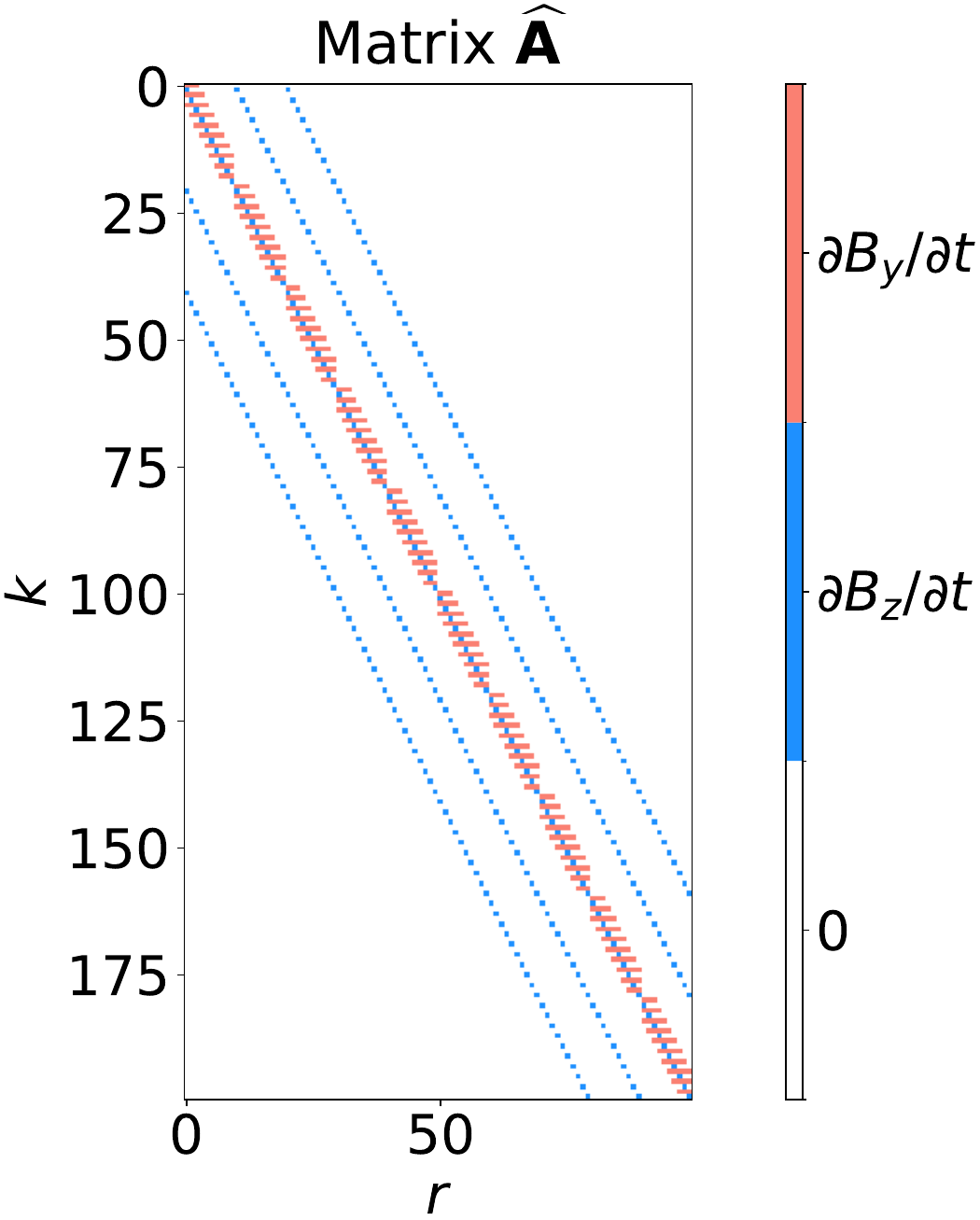}
    \caption{Structure of the matrix of the system $\widehat{{\bf A}} $ for a $N \times M = 10 \times 10$ domain defined by $k$ rows and $r$ columns. In red we show the position of the $\widehat{A}_{kr}$ coefficients related to Eq.~\ref{eq:induction_eq_2D_yz_y} and in blue those related to Eq.~\ref{eq:induction_eq_2D_yz_z}.}
    \label{fig:matrix}
\end{figure}

In matrix form, the system of equations can be written as $\widehat{{\bf A}} \cdot {\bf x} = {\bf c}$, where $\widehat{{\bf A}}$ is constructed from the known coefficients $a_{i,j}$ and $b_{i,j}$. This matrix has dimensions $2 \cdot N \cdot M \times N \cdot M$ for a domain of $N \times M$ grid points. ${\bf x}$ is the vector of unknowns with dimensions $N\cdot M \times 1$, while ${\bf c}$ corresponds to the vector of independent coefficients ($c_{i,j}^a$, $c_{i,j}^b$) with dimensions $2 \cdot N \cdot M \times 1$. In the matrix $\widehat{{\bf A}}$, we will have two rows (one for each available component of the induction equation used) associated with the same unknown $v_{y}(i,j)$. The following piece of pseudo-code can be used to build the $\widehat{A}_{kr}$ elements of $\widehat{{\bf A}}$ as well as the $c_k$ elements of the vector $\ve{c}$:

\begin{algorithmic}[1]
    \State $r = 0$
    \For {$k=0,2\cdot N\cdot M-1,2$}
    \newline
        \Comment{Coordinates of the grid point $(i,j)$ are given by}
        \State $ i = \textrm{int}(r/M)$ 
        \State $j = \textrm{mod}(r,M)$
        \newline
    \Comment{Elements referring to $\dot{B_y}$ (Eq.~\ref{eq:induction_eq_2D_yz_y})}
                \begin{eqnarray}\label{eq:bydot_coefs}
                A(k, i\cdot M + j-2) & = & a_{i,j}^{i,j-2} \notag\\
                A(k, i\cdot M + j-1) & = & a_{i,j}^{i,j-1} \notag\\
                A(k, i\cdot M + j) & = & a_{i,j}^{i,j}\\
                A(k, i\cdot M + j+1) & = & a_{i,j}^{i,j+1} \notag\\
                A(k, i\cdot M + j+2) & = & a_{i,j}^{i,j+2} \notag\\
                c(k) & = & c^{a}_{i,j}\notag
                \end{eqnarray}
    \newline
    \Comment{Elements referring to $\dot{B_z}$ (Eq.~\ref{eq:induction_eq_2D_yz_z})}
                \begin{eqnarray}\label{eq:bzdot_coefs}
                A(k+1, (i-2)\cdot M + j) & = & b_{i,j}^{i-2,j} \notag\\
                A(k+1, (i-1)\cdot M + j) & = & b_{i,j}^{i-1,j} \notag\\
                A(k+1, i\cdot M + j) & = & b_{i,j}^{i,j}\\
                A(k+1, (i+1)\cdot M + j) & = & b_{i,j}^{i+1,j} \notag\\
                A(k+1, (i+2)\cdot M + j) & = & b_{i,j}^{i+2,j} \notag\\
                c(k+1) & = & c^{b}_{i,j} \notag
                \end{eqnarray}
    \State $r = r+1$
    \EndFor
\end{algorithmic}

\noindent where the for-loop indexes refer to the starting value, end value, and step value. An example of the $\widehat{{\bf A}}$ matrix for a $10 \times 10$ domain is shown in Fig.~\ref{fig:matrix}. As explained above, for this particular example, $\widehat{{\bf A}}$ has a size of $200 \times 100$ (i.e., 200 equations for 100 unknowns). The color code indicates whether that particular point on the matrix comes from Eq.~\ref{eq:induction_eq_2D_yz_y} (red) or Eq.~\ref{eq:induction_eq_2D_yz_z} (blue).

\subsection{Boundary conditions}
\label{subsection:boundary_conditions}

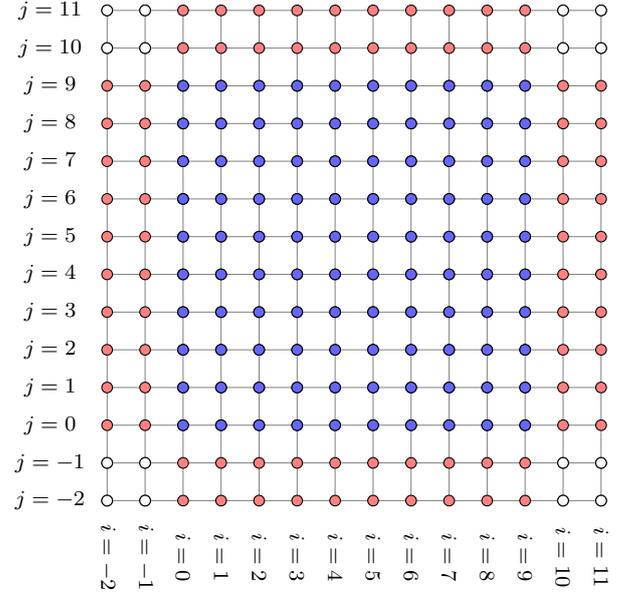
\begin{figure}[h]
\centering
\begin{tikzpicture}[scale=0.5]
    \draw[step=1cm,gray,very thin] (-2,0) grid (11,13);
    
    \node at (-3.5,13) {\tiny $j=11$};
    \node at (-3.5,12) {\tiny $j=10$};
    \node at (-3.5,11) {\tiny $j=9$};
    \node at (-3.5,10) {\tiny $j=8$};
    \node at (-3.5,9) {\tiny $j=7$};
    \node at (-3.5,8) {\tiny $j=6$};
    \node at (-3.5,7) {\tiny $j=5$};
    \node at (-3.5,6) {\tiny $j=4$};
    \node at (-3.5,5) {\tiny $j=3$};
    \node at (-3.5,4) {\tiny $j=2$};
    \node at (-3.5,3) {\tiny $j=1$};
    \node at (-3.5,2) {\tiny $j=0$};
    \node at (-3.5,1) {\tiny $j=-1$};
    \node at (-3.5,0) {\tiny $j=-2$};
    
    \foreach \x/\label in {-2/-2,-1/-1,0/0,1/1,2/2,3/3,4/4,5/5,6/6,7/7, 8/8, 9/9, 10/10, 11/11} {
        \node[rotate=270] at (\x, -1.5) {\tiny $i=\label$};
    }

    \foreach \x in {-2,...,11}
    \foreach \y in {0,...,13}
    {
    \node[circle,draw=black, fill=red!50, inner sep=0pt,minimum size=4pt] at (\x,\y) {};
    }

    \foreach \x in {0,...,9}
    \foreach \y in {2,...,11}
    {
    \node[circle,draw=black, fill=blue!60, inner sep=0pt,minimum size=4pt] at (\x,\y) {};
    }

    \foreach \x/\y in {-2/0, -2/1, -1/0, -1/1, -2/12, -2/13, -1/12, -1/13, 10/0, 11/0, 10/1, 11/1 , 10/12, 11/13, 11/12, 10/13}
    {
    \node[circle,draw=black, fill=white, inner sep=0pt,minimum size=4pt] at (\x,\y) {};
    }
\end{tikzpicture}
  \caption{Example of a 2D grid for a $10 \times 10$ domain. The solution of the system is evaluated in the blue points. These points are called "inner cells". The boundary condition must be specified in all red points. These are "ghost" or "boundary" cells. The white points in the corners are not used in the scheme considered here. In a different numerical scheme, they may play a role.}
  \label{fig:2D_grid}
\end{figure}

To define the boundary conditions while maintaining the derivative scheme given by Eq.~\ref{eq:first_derivative_5_point} throughout the entire domain, we introduced "ghost cells" around the boundaries of the domain (see Fig.~\ref{fig:2D_grid}). Although ghost cells are not part of the domain itself, we assumed a certain behavior of the unknown $v_y$ function at those points. This behavior implies changes to the elements of the matrix $\widehat{\bf A}$ and to the components of the vector $\ve{c}$ related to the points close to the boundary. Since we used a five-point stencil scheme for the derivatives, the grid points whose coefficients will be modified are those with $i = 0, 1, N-2, N-1$ and/or $j=0, 1, M-2, M-1$ (see Fig.~\ref{fig:2D_grid}), which refer to grid cells at the corners or sides of the domain, respectively.

\noindent How the coefficients $\widehat{A}_{rk}$ of matrix $\widehat{\bf{A}}$ and the $c_k$ components of vector $\ve{c}$ are modified close to the boundary depends on the actual boundary conditions employed. If we assume Dirichlet boundary conditions, we must set the value of $v_y$ at the boundary. For example, $v_{y} = f_{D}$, where $f_{D}$ denotes a fixed value and the subscript $D$ indicates that Dirichlet boundary conditions are being applied. Consequently, the values of $v_y$ in the ghost cells are no longer unknowns, and the terms of Eqs.~\ref{eq:induction_eq_2D_yz_y_5p_simpl} and~\ref{eq:induction_eq_2D_yz_z_5p_simpl} that involve $v_y$ at the ghost cells become independent coefficients. For that reason, we get new $\ve{c}$ coefficients (referred to as $c^{a,\star}$ and $c^{b,\star}$) that can be written as a function of the old coefficients  ($c^{a}$ and $c^{b}$; given by Eqs.~\ref{eq:ac_coeffs} and \ref{eq:bc_coeffs}) as

\begin{eqnarray}
\label{eq:indep_coeff_yz_00}
c^{a,\star}_{0,0} & = & c^{a}_{0,0} - a_{0,0}^{0,-2} \cdot f_{D}(0,-2) - a_{0,0}^{0,-1} \cdot f_{D}(0,-1) \\
c^{b,\star}_{0,0} & = & c^{b}_{0,0}  - b_{0,0}^{-2,0} \cdot f_{D}(-2,0) -b_{0,0}^{-1,0} \cdot f_{D}(-1,0).
\end{eqnarray}

We note that in the case of Dirichlet boundary conditions, the $\widehat{A}_{rk}$ elements of matrix $\widehat{\bf{A}}$, built from $a_{i,j}$ and $b_{i,j}$ according to Eqs.~\ref{eq:bydot_coefs} and~\ref{eq:bzdot_coefs}, do not change.

On the other hand, we can assume symmetric boundary conditions by defining  $f_{-1} = f_{0}$ and $f_{-2} = f_{1}$. These boundary conditions also ensure that, at some point in between the domain and the ghost zone, the derivative of $v_y$ is zero (Neumann boundary condition). Unlike the case of the Dirichlet boundary conditions, where $v_y=f_D$ was assumed to be known, Neumann boundary conditions link $v_y$ in the ghost cells to the unknown $v_y$ inside the domain. For this reason, close to the boundary we must modify the related $\widehat{A}_{rk}$ elements of matrix $\widehat{\ve{A}}$ given by Eqs.~\ref{eq:ac_coeffs} and ~\ref{eq:bc_coeffs}. The new coefficients, $a^{\star}$ and $b^{\star}$, derived from this assumption can be written as a function of the old $a$ and $b$ coefficients (given by Eqs.~\ref{eq:bydot_coefs} and~\ref{eq:bzdot_coefs}). Here we provide examples for a corner point and a boundary line:

\begin{eqnarray}
\left.\begin{tabular}{cc}
$a^{0, 0, \star}_{0,0} =$ & $a^{0, 0}_{0,0} + a_{0,0}^{0,-1}$ \\
$a^{0, 1,\star}_{0,0} =$ & $a^{0, 1}_{0,0} + a_{0,0}^{0,-2}$ \\
$b^{0,0,\star}_{0,0} =$  & $b^{0,0}_{0,0} + b_{0,0}^{-1,0}$ \\
$b^{1,0,\star}_{0,0} =$ & $b^{1, 0}_{0,0} + b_{0,0}^{-2,0}$
\end{tabular}\right\} \;\; \textrm{if $(i,j) = (0,0)$} \label{eq:coef_yz_00}\\
\left.\begin{tabular}{cc}
$b^{0,j,\star}_{0,j} =$ & $b^{0,j}_{0,j} + b_{0,j}^{-1,j}$ \\
$b^{1,j,\star}_{0,j} =$ & $b^{1, j}_{0,j} + b_{0,j}^{-2,j}$ 
\end{tabular}\right\} \;\; \textrm{if $(i,j) = (0,j)$} \label{eq:coef_yz_0j}\\
\left.\begin{tabular}{cc}
$b^{0, j, \star}_{1,j} =$ & $b^{0,j}_{1,j} + b_{1,j}^{-1,j}$ 
\end{tabular}\right\} \;\; \textrm{if $(i,j) = (1,j)$} \label{eq:coef_yz_1j},
\end{eqnarray}

\noindent where $ 1 < j < n-2$ in Eqs.~\ref{eq:coef_yz_0j},~\ref{eq:coef_yz_1j}. Finally, we note that in the case of Neumann boundary conditions, the $c_k$ coefficients of $\ve{c}$ remain the same.

\newpage

\subsection{Solution of the linear system}
\label{subsection:linear_system_solution}

Having defined the matrix of the system $\widehat{\bf{A}}$, the vector of independent coefficients $\ve{c}$, and the boundary conditions, we could then solve the linear system of equations and infer the unknown horizontal velocity field $v_{y}(y,z)$ that is encoded in the vector $\ve{x}$. Owing to the fact that we have an overdetermined system with $2\cdot N \cdot M$ equations and $N\cdot M$ unknowns, we solved the linear system in a least square sense: $\ve{x} = (\widehat{\bf{A}}^\intercal \widehat{\bf{A}})^{-1} \widehat{\bf A}^\intercal \ve{c}$, where the "$^\intercal$" symbol indicates the transpose of the matrix. The solution of the linear system is found through the well-known LAPACK\footnote{\url{https://netlib.org/lapack/}} package, in particular the subroutine dgesv that uses LU decomposition for a square matrix.

\section{Comparison between analytical and numerical solutions}
\label{section:comparison_analytical_numerical}

\subsection{Two-dimensional analytical magnetic and velocity fields}
\label{subsection:no_temporal_yz}

\begin{figure}[h!]
    \centering
    \includegraphics[width=\linewidth]{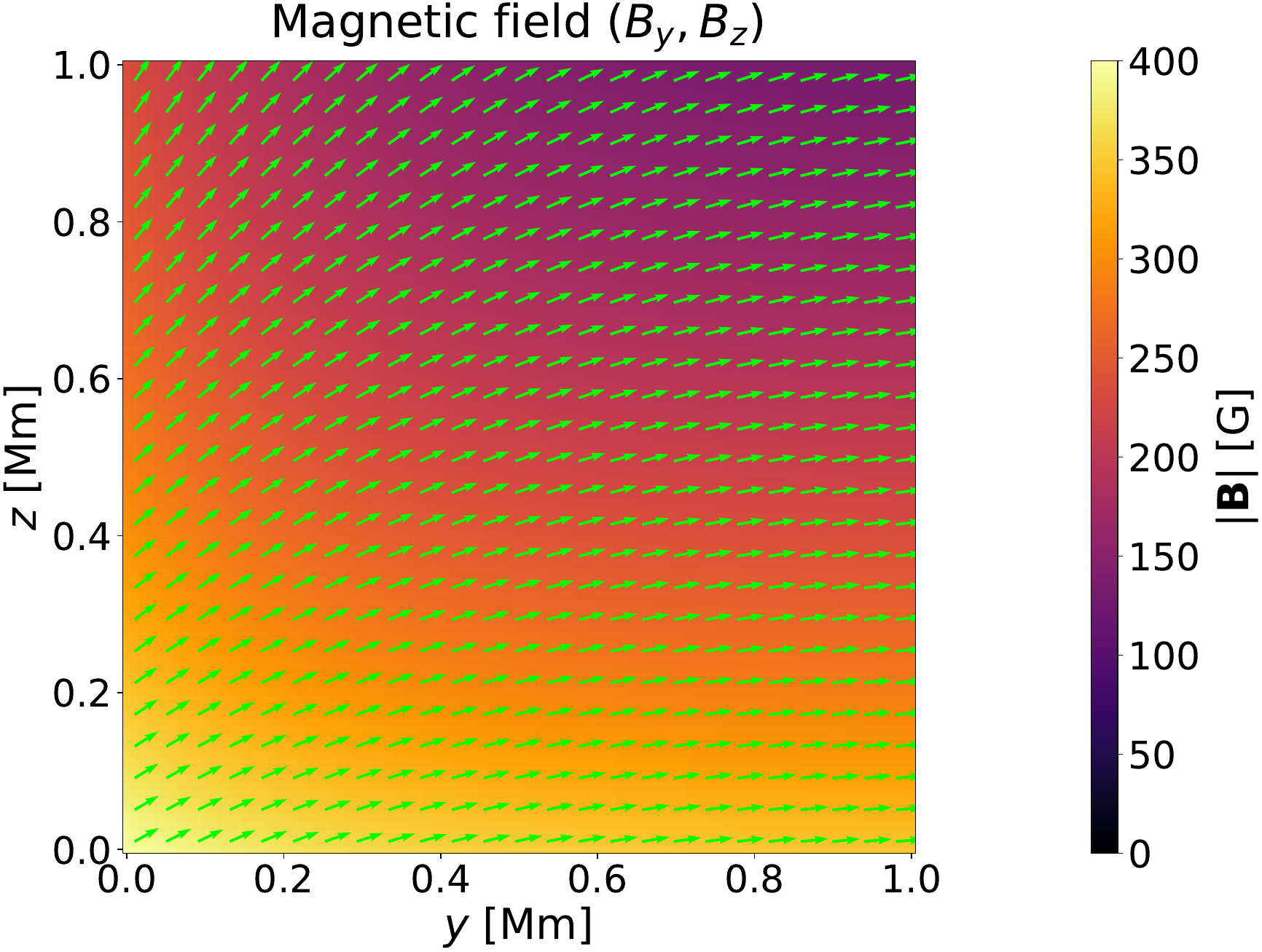}
    \caption{Two-dimensional analytical magnetic field. The color map shows the magnitude of the magnetic field $\|{\bf B}\|$. The green arrows show its direction. The color bar is saturated at $400$ G to show the lower values of the magnetic field. The maximum value of the magnetic field at the selected region is $\approx 415$ G.}
    \label{fig:B_2D}
\end{figure}

We defined a two-dimensional domain in the $(y,z)$ plane. This domain is discretized in $N \times M = 100 \times 100$ grid cells. Each grid cell is considered to be squared $h_y = h_z =h$ with a length of $h=10^6$~cm. In this domain, we prescribe the following components for the magnetic and velocity fields at instant $t_0$:

\begin{eqnarray}
    B_{y} & = & B_{1} \exp{\frac{\alpha}{\lambda}z} 
    \label{eq:yz_By_ex}\\
    B_{z} & = & B_{2} \exp{\frac{\alpha}{\lambda}y} 
    \label{eq:yz_Bz_ex}\\
    v_{y} & = & v_1 \cos\left\{\frac{2 \pi (My - Nz)}{NMh}\right\} \
    \label{eq:yz_vy_ex}\\
    v_{z} & = & v_1 \frac{M}{N} \cos\left\{\frac{2 \pi (My - Nz)}{hNM}\right\} + v_2 \cos\left\{\frac{2 \pi y}{hN}\right\},
    \label{eq:yz_vz_ex}
\end{eqnarray}

\noindent where $B_{1} = 350$ Gauss, $B_{2} = 200$ Gauss, $\alpha = -0.1$, $\lambda = 10h$, $v_1 = 10^5$ cm~s$^{-1}$, and $v_2 = v_1 / 2$. We note that since $B_y$ does not depend on $y$ and $B_z$ does not depend on $z$, the solenoidal condition for the magnetic field, $\nabla \cdot \ve{B}=0$, is trivially satisfied. The detailed configuration of the magnetic field prescribed above is presented in Fig.~\ref{fig:B_2D}.

Equations~\ref{eq:yz_By_ex}-\ref{eq:yz_vz_ex} were then applied to Equations~\ref{eq:induction_eq_2D_yz_y}-\ref{eq:induction_eq_2D_yz_z} to find the analytical expressions for $\dot{B_y}$ and $\dot{B_z}$. In the following, we assumed $\dot{B_y}$, $\dot{B_z}$, $B_z$, $B_y$, and $v_z$ to be known, leaving the horizontal component of the velocity as the only unknown, which we refer to as $\widetilde{v_y}(y,z)$. To determine $\widetilde{v_{y}}$, we employed vanishing Neumann boundary conditions instead of Dirichlet boundary conditions (see Sect.~\ref{subsection:boundary_conditions}) because the latter require a knowledge of $v_y$ at the boundaries, which in practical applications we do not have. To this end, we employed the expressions in Eqs.~\ref{eq:coef_yz_00} -~\ref{eq:coef_yz_1j} extended to all the points close to the boundary (i.e., $i=0,1,N-2,N-1$ and $j=0,1,M-2,M-1$ for an $N \times M$ domain based on the indexing from Fig.~\ref{fig:2D_grid}).

\begin{figure*}[h!]
    \centering
    \includegraphics[width=0.9\linewidth]{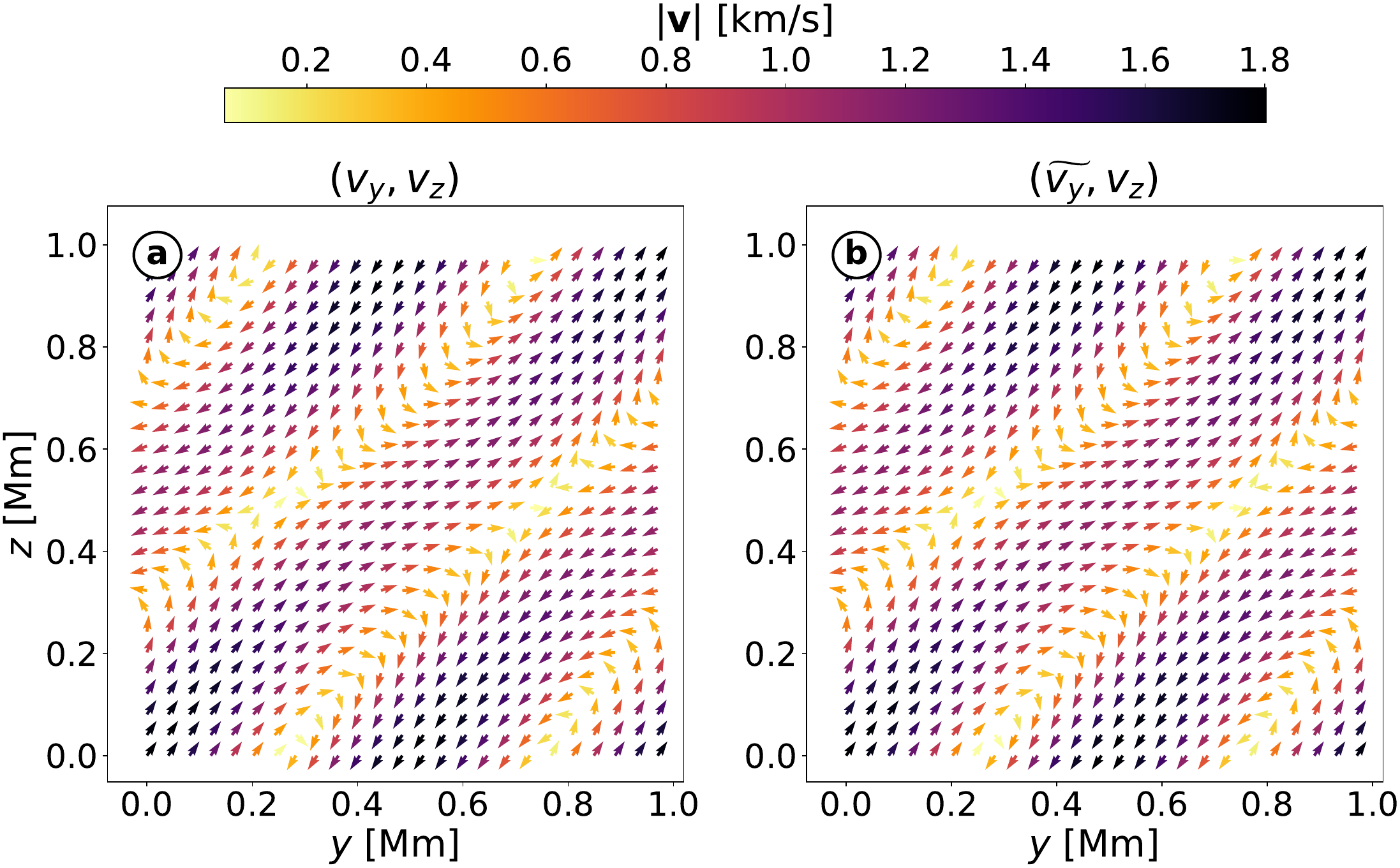}
    \caption{Velocity field maps in the plane $(y,z)$. Panel (a) shows the velocity field defined analytically: $\ve{v} = (v_y,v_z)$ (see Eqs.~\ref{eq:yz_vy_ex} and ~\ref{eq:yz_vz_ex}), whereas panel (b) provides the inferred velocity field: $\widetilde{\ve{v}} = (\widetilde{v_y},v_z)$. The modulus of the vectors is encoded in the color of the arrows as indicated in the color bar.}
    \label{fig:yz_Neu_ex1}
\end{figure*}

With this, we applied the approach described in Sect.~\ref{section:numerical_method} to retrieve $\widetilde{v_y}(y,z)$, which we then compared with the original $v_y(y,z)$ given by Equation~\ref{eq:yz_vy_ex}. Figure~\ref{fig:yz_Neu_ex1} shows the analytical velocity field $\ve{v} = (v_y,v_z)$ defined through Eq.~\ref{eq:yz_vy_ex} and Eq.~\ref{eq:yz_vz_ex} (panel a) and the inferred velocity field $\ve{\widetilde{v}} = (\widetilde{v_y},v_z)$ (panel b). Although at first glance it seems that both velocities are very similar, in the sense of direction and magnitude, one must bear in mind that these plots do not provide a complete picture because $v_z$ is the same in both panels (i.e., it was assumed to be known).

\begin{figure*}[h!]
    \centering
    \includegraphics[width=0.9\linewidth]{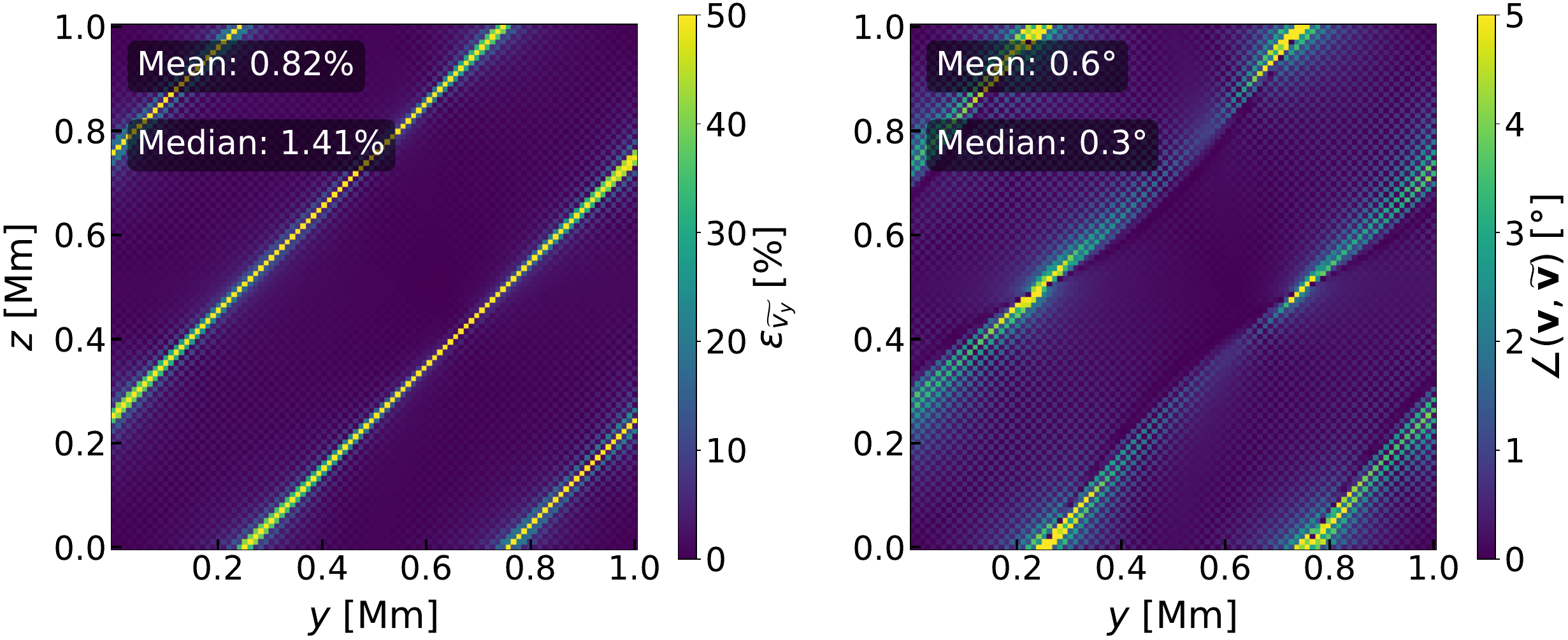}
    \caption{Errors of the inference of $\widetilde{v_{y}}$ in the analytical 2D case. The left panel shows the relative error, $\varepsilon_{\widetilde{v_{y}}}$, of the inferred velocity. The right panel shows the angle between the analytical velocity (${\bf v}$) and the inferred one (${\widetilde{\bf v}}$). See text for more details. Both plots also indicate the mean and the median of these errors.}
    \label{fig:yz_errors_Neu_ex1}
\end{figure*}

To further investigate the accuracy with which our method can determine the horizontal component of the velocity field in this analytical two-dimensional case, we show in Fig.~\ref{fig:yz_errors_Neu_ex1} the relative error of the inferred velocity, defined as $\varepsilon_{\widetilde{v_{y}}} = |\frac{\widetilde{v_{y}} - v_{y}}{v_{y}}|$ (left panel), and the angle formed by the analytical and the inferred velocity fields, defined as $\angle(\ve{v},\widetilde{\ve{v}}) = \arccos(\ve{v} \cdot \widetilde{\ve{v}} / [\|\ve{v}\| \|\widetilde{\ve{v}} \|])$ (right panel).

As can be seen, the average and median errors, $\varepsilon_{\widetilde{v_{y}}}$, over the entire domain are less than 1\% and 1.5\%, respectively. In addition, the angles between the original velocity field and the retrieved one, $\angle(\ve{v},\widetilde{\ve{v}})$, are also very small -- about $0.5^{\circ} $ on average. Over the majority of the domain, the actual retrieval is much better: $\varepsilon_{\widetilde{v_{y}}} \le 0.01\%$.

\subsection{CO5BOLD numerical simulation in (y,z)}
\label{subsection:simulation_yz}

The analytical test in Sect.~\ref{subsection:no_temporal_yz} features relatively simple magnetic and velocity fields (see e.g., Fig.~\ref{fig:B_2D}). To test our method with a more realistic scenario, in this section we perform a similar test but using magnetic $\ve{B}(t_0)$ and velocity $\ve{v}(t_0)$ fields from a two-dimensional MHD numerical simulation of solar surface magneto-convection obtained from the CO5BOLD code \citep{freytag2012simulations}. The aforementioned simulation provides us with $B_{y}(y,z), B_{z}(y,z), v_{y}(y,z)$, and $v_{z}(y,z)$ at a fixed time step, $t_0$. In addition, the simulation has vanishing $B_{x}$ and $v_{x}$. Details on how the 2D simulation was carried out are provided in Appendix~\ref{sec:cobold}.

An important point to mention here pertains to the magnetic diffusivity, $\eta$. Although CO5BOLD solves the induction equation under ideal MHD, numerical diffusion is unavoidable, and therefore, if we were to take the magnetic and velocity fields at two different time steps, $t_{+}$ and $t_{-}$, separated by a small $\Delta t = t_{+} - t_{-}$ and centered around $t_0$, the ideal induction equation would not be satisfied:

\begin{equation}
\dot{\ve{B}}(t_0) = \lim_{\Delta t \rightarrow 0} \frac{\ve{B}(t_{+})-\ve{B}(t_{-})}{2 \Delta t} \ne \nabla \times [\ve{v}(t_0) \times \ve{B}(t_0)].
\label{eq:non_ideal_induction}
\end{equation}

For this reason, we do not evaluate the time derivative at $t_0$ from two different snapshots ($t_{+}$ and $t_{-}$) of the MHD simulation. Instead we determine $\dot{\ve{B}}(t_0)$ directly from $\ve{v}(t_0)$ and $\ve{B}(t_0)$ using the right-hand side of Eq.~\ref{eq:induction_eq_vect_eval}. This ensures that the induction equation in ideal MHD is satisfied. This step can be justified because at this point we are only interested in studying the performance of the method described in Sect.~\ref{section:numerical_method} under ideal conditions. We defer the investigation of the effects of numerical or magnetic diffusivity, $\eta_{\rm eff}$ (see Appendix~\ref{sec:cobold}), for a future study, which should also include the effect of having a finite $\Delta t$ such that the limit in Eq.~\ref{eq:non_ideal_induction} is only approximately satisfied.

\begin{figure}[h!]
    \centering
    \includegraphics[width=\linewidth]{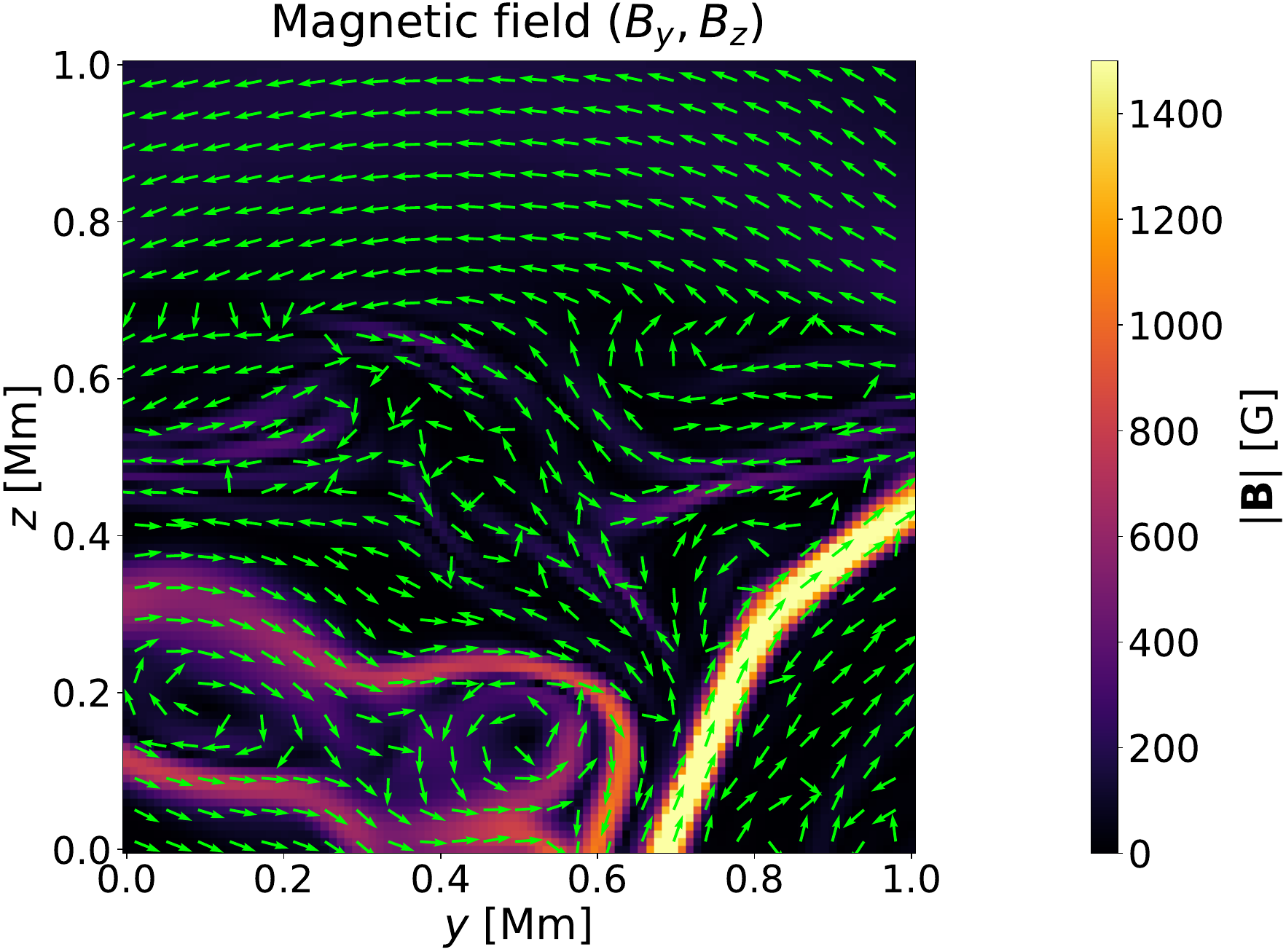}
    \caption{Two-dimensional CO5BOLD simulation magnetic field. The color map shows the magnitude of the magnetic field $\|\ve{B}\|$. The green arrows show its direction. The color bar is saturated at $1500$ G to show the lower values of the magnetic field. The maximum value of the magnetic field within the selected region is $\approx 2245$ G.}
    \label{fig:B_sim_2D}
\end{figure}

\begin{figure*}[h!]
    \sidecaption
    \includegraphics[width=0.75\linewidth]{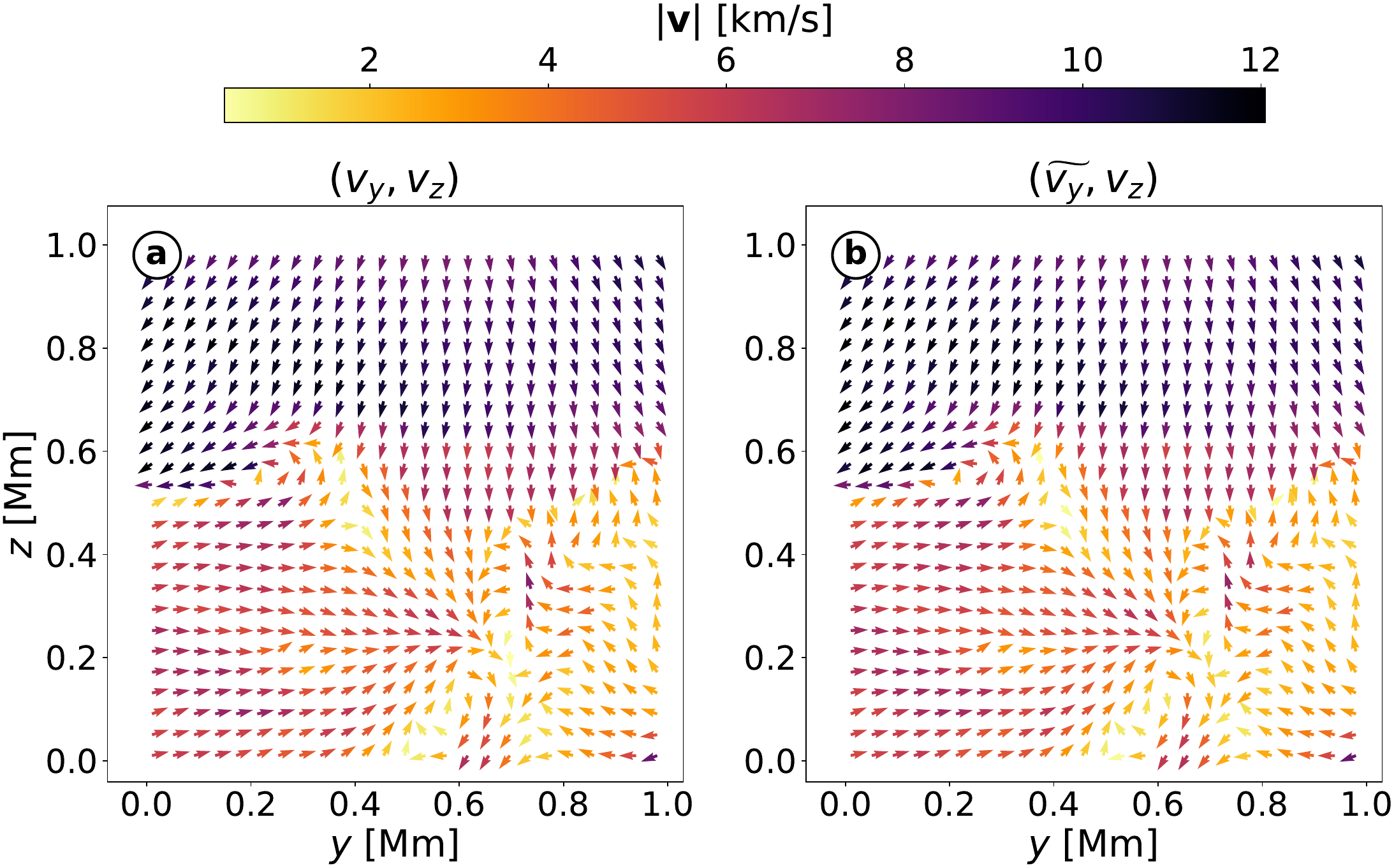}
    \caption{Same as Fig.~\ref{fig:yz_Neu_ex1} but for the case of the two-dimensional MHD numerical simulation.}
    \label{fig:yz_sim_2D}
\end{figure*}

Next, we took a central region of $N \times M =100 \times 100$ grid points from the full domain of the simulation and determined the horizontal component of the velocity $v_y(y,z)$ from the known $\dot{\ve{B}}$, $\ve{B}$, $v_z$ and by employing vanishing Neumann boundary conditions. Figure~\ref{fig:B_sim_2D} illustrates the magnetic field within this region, whereas Fig.~\ref{fig:yz_sim_2D} compares the original and the retrieved velocity fields. This last figure shows that the inferred velocity field is perceptually identical to the original from the MHD simulation.

\begin{figure*}[h!]
    \sidecaption
    \includegraphics[width=0.75\linewidth]{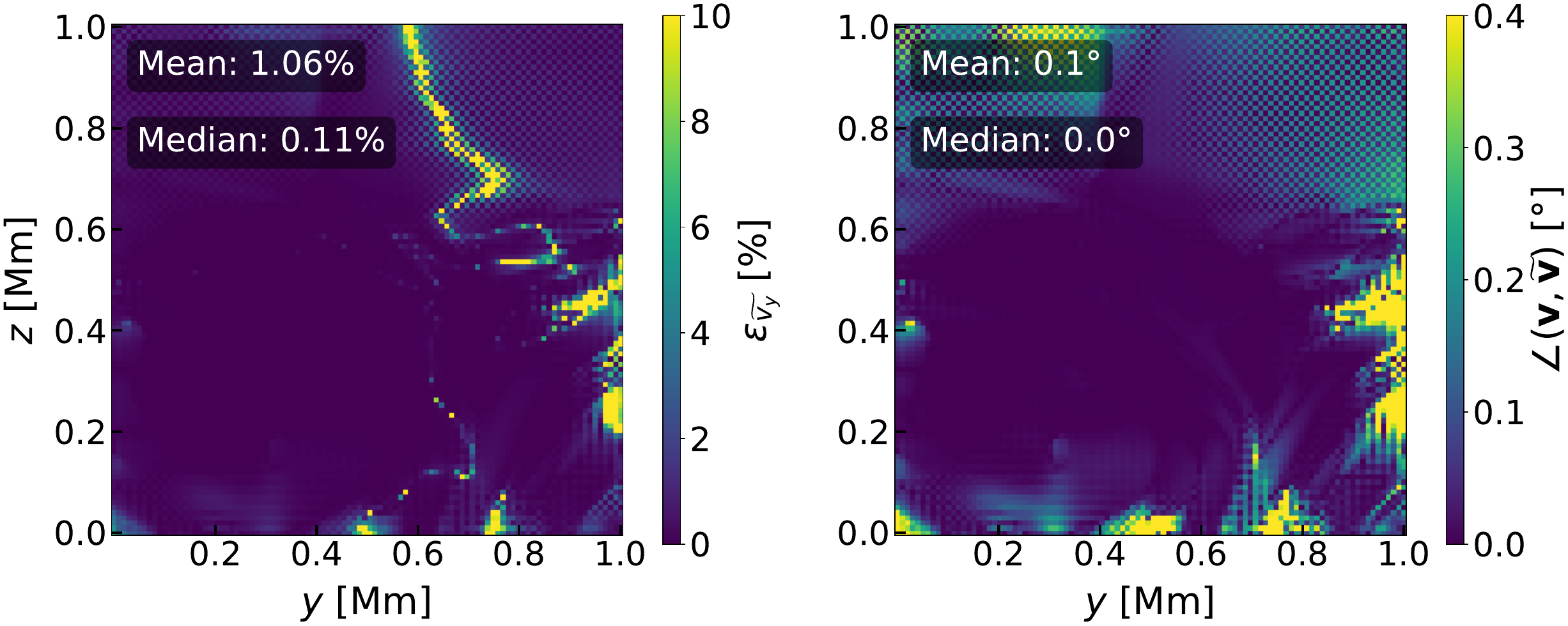}
    \caption{Same as Fig.~\ref{fig:yz_errors_Neu_ex1} but for the case of the two-dimensional MHD simulation.}
    \label{fig:yz_errors_sim_2D}
\end{figure*}

A more detailed study of the errors is presented in Fig.~\ref{fig:yz_errors_sim_2D}, where again the error in the inference of $v_y$, $\varepsilon_{\widetilde{v_{y}}}$ and the errors in the inference of the angle of the velocity field, $\angle(\ve{v},\widetilde{\ve{v}})$, are presented on the left and right panels, respectively. As can be seen, now the mean and median values of the errors are smaller than in the 2D analytical case (Sect.~\ref{subsection:no_temporal_yz}).

\section{Discussion}
\label{section:discussion}

\subsection{Errors}
\label{subsection:errors}

In the previous sections, we showed that the method developed in this work retrieves $v_y(y,z)$ with a very high accuracy: about 1\% mean error in the inference of $v_y$ and less than 0.5$^{\circ}$ in the inference of the orientation of the total velocity field. We have referred to these two quantities as $\varepsilon_{\widetilde{v_{y}}}$ and  $\angle(\ve{v},\widetilde{\ve{v}})$, respectively.

Interestingly, there are some specific regions where the values of $\varepsilon_{\widetilde{v_{y}}}$ and $\angle(\ve{v},\widetilde{\ve{v}})$ are much larger than the mean and median values. For instance, this is the case of the diagonal stripes in Fig.~\ref{fig:yz_errors_Neu_ex1} (analytical case), where $\varepsilon_{\widetilde{v_{y}}}$ and $\angle(\ve{v},\widetilde{\ve{v}})$ reach values as high as 50\% and 5$^{\circ}$, respectively. Similar regions of large uncertainties also appear in Fig.~\ref{fig:yz_errors_sim_2D} for the case of the 2D MHD simulation, albeit with smaller errors than in the analytical case.

To investigate the source of the large uncertainties in these regions, it is convenient to notice that, as per their definitions,  $\varepsilon_{\widetilde{v_{y}}}$ will always be larger in those regions where $v_y \rightarrow 0$, whereas $\angle(\ve{v},\widetilde{\ve{v}})$ will be larger in regions where $\|\ve{v}\| \rightarrow 0$. It is also important to note that $\|\ve{v}\| \rightarrow 0$ immediately implies $v_y \rightarrow 0$, but not the other way around.

With this clarified, we are in a position to assert that in the analytical case (Sect.~\ref{subsection:no_temporal_yz}), $\varepsilon_{\widetilde{v_{y}}}$ and $\angle(\ve{v},\widetilde{\ve{v}})$ are the largest in regions where $\|\ve{v}\| \rightarrow 0$ (see Fig.~\ref{fig:yz_Neu_ex1}). In the case of the two-dimensional MHD simulation (Sect.~\ref{subsection:simulation_yz}), the regions where the errors are largest do not coincide with locations where $\|\ve{v}\| \rightarrow 0$. We can ascertain this because in the MHD simulations, the minimum velocity is about 1 km~s$^{-1}$ (see color bar in Fig.~\ref{fig:yz_sim_2D}). In the regions where $v_y \rightarrow 0$, and thus where the velocity is mostly aligned with the $z$-axis, $\varepsilon_{\widetilde{v_{y}}}$ instead increases. Also, the fact that the modulus of the velocity in the MHD simulations is never zero implies that the angle between the velocity vector in the MHD simulations and inferred velocity vector, $\angle(\ve{v},\widetilde{\ve{v}})$, does not increase as much as in the analytical example presented in Sect.~\ref{subsection:no_temporal_yz}.

\begin{figure}[h!]
    \centering
    \includegraphics[width=\linewidth]{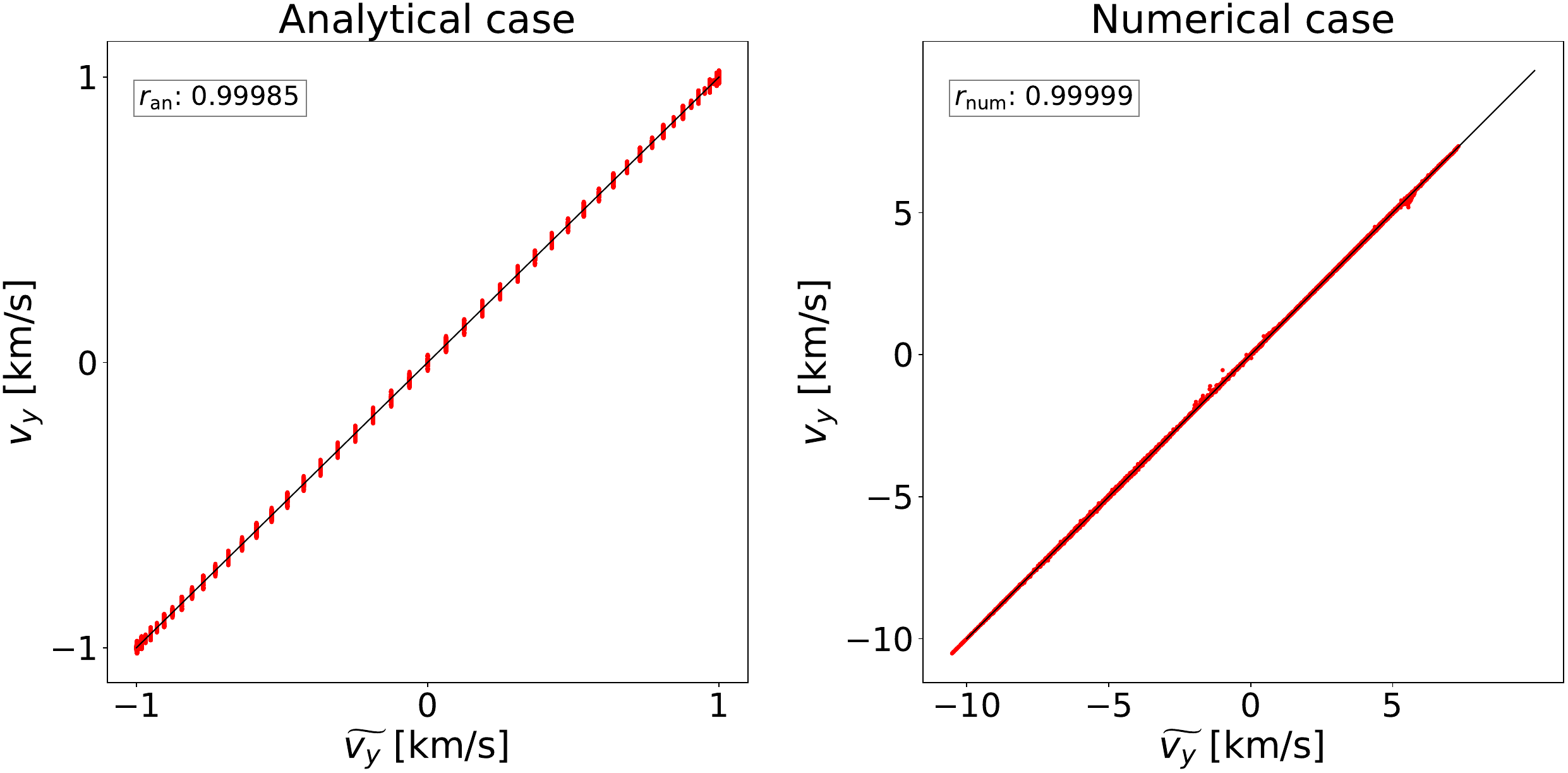}
    \caption{Scatter plots of $v_{y}$ and $\widetilde{v_{y}}$ for the analytical (left) and MHD numerical simulation (right) cases. The Pearson correlation coefficient is measured in both cases ($r$). The slope of one line (indicating a perfect retrieval) is displayed as a black line.}
    \label{fig:pearson_corr}
\end{figure}

To showcase how well the horizontal component of the velocity $v_y$ is inferred in a way that is independent from the magnitude of $v_y$ or $\|{\bf v}\|$, we show in Fig.~\ref{fig:pearson_corr} scatter plots between the real $v_y$ and the inferred values of the velocity $\widetilde{v_{y}}$. One can see that in the analytical test (left; Sect.~\ref{subsection:no_temporal_yz}) and the test with MHD simulations (right; Sect.~\ref{subsection:simulation_yz}), the Pearson correlation coefficient ($r$) is almost 1.0. Furthermore, almost all points of the 2D domain fall almost exactly on top of the perfect retrieval line (i.e., slope one) indicated by the black line.

\subsection{Known degeneracies}
\label{subsection:degeneracies}

Due to the structure of the induction equation, there are theoretically two degeneracies that can affect the retrieval of the velocity vector. First, it is possible to add a gradient of a scalar function, $\phi$, to the cross product $\ve{v} \times \ve{B}$ without affecting $\ve{\dot{B}}$. Second, only the velocity component perpendicular to the magnetic field contributes to its variation over time.\\

These two degeneracies suggest that the velocity field inferred using the method presented in this paper may not represent the real velocity of the plasma. This raises the question of how we can ensure that the retrieved velocity is the one we seek.\\

In two dimensions, neither degeneracy is very relevant. For instance, in the $(y,z)$ plane, we define $\ve{v}$ and $\ve{B}$ both perpendicular to the $x$-axis. Then, $\ve{v} \times \ve{B}$ is parallel to the unit vector $\ve{e}_{x}$. Moreover, if we consider a scalar function, $\phi$, whose gradient can be added to the cross product, $\ve{v} \times \ve{B}$, then by definition, $\nabla \phi$ is also parallel to $\ve{e}_{x}$. However, given that $\phi$ is only a function of $(y, z)$, it trivially follows that $\nabla \phi = \frac{\partial \phi(y,z)}{\partial x} \ve{e}_{x} = 0$.\\

In three dimensions, the two aforementioned degeneracies can potentially play a much more important role. We anticipate being able to retrieve the real plasma velocity field in 3D for two reasons. First,  $v_{z}$ is considered to be known (i.e., from the Stokes inversion that measures it from the Doppler effect), and this component of the velocity already includes contributions from the parallel and perpendicular components of the velocity with respect to $\ve{B}$. For this reason, the knowledge of $v_{z}$ breaks the degeneracy, and it allows us to retrieve the full velocity field through the induction equation. Second, our method can be generalized to include other MHD constraints, such as mass conservation. These additional constraints could help us obtain the plasma velocity field independently of the term $\nabla \phi$ that can be added to $\ve{v} \times \ve{B}$. We will address these topics in more detail in future work.

\section{Conclusions and future work}
\label{section:conclusions}

We have developed a new method to infer the velocities perpendicular to the line of sight in the solar atmosphere. This method employs the full induction equation in ideal MHD and assumes that the line-of-sight component of the velocity, $v_{z}$; the magnetic field, $\ve{B}$; and its time derivative, $\dot{\ve{B}}$, are known as a function of space and time. Nowadays, this can be achieved thanks to novel Stokes inversion techniques applied to data from modern spectropolarimeters \citep{anjali2020ifu,liu2025mihi}.

As a consequence of using the three components of the induction equation, the method also allows retrieval of the full $z$ dependence of the horizontal velocity. This is not possible when employing only the vertical ($z$ direction) component of the induction equation \citep[see, e.g.,][]{longcope2004induction}.

As a first step, we tested our newly developed method in two-dimensions by first using analytically prescribed functions for the velocity and magnetic fields and secondly by using velocity and magnetic fields from MHD simulations of the solar surface magnetoconvection. In these 2D cases, only two of the components of the induction equation were used. Our results indicate that in an ideal scenario, it is possible to retrieve the horizontal component of the velocity with an average error of about 1\%. In regions where the horizontal component of the velocity is very small, uncertainties in the direction and relative error in the magnitude become significantly larger. Overall, Pearson's correlation coefficient is very close to unity ($r \simeq 1$) in both of the studied cases. This allows us to conclude that our method can retrieve the velocity correctly.

An important feature of the method presented in this paper is that it can be easily extended to three dimensions by including the third component of the induction equation. Although the discretization of the equations and the construction of the matrix $\widehat{\bf A}$ follow the same procedure described in this paper (see Eqs.~\ref{eq:bydot_coefs}-\ref{eq:bzdot_coefs}), the matrix dimensions increase substantially as the domain expands from two to three dimensions. Specifically, the size of the matrix, $\widehat{\bf A}$,  changes from $2NM \times NM$ to $3NML \times 2NML$ because the system of equations includes an additional dimension as well as an additional equation with two unknowns ($v_x$,$v_y$) at each point of the domain. Inverting a matrix of this size requires a considerable amount of computational time, but we plan on undertaking this step next. Apart from extending our approach to three dimensions, in the future we hope to investigate the accuracy with which we can infer the horizontal components of the velocity under more realistic scenarios. Namely, we plan to examine the presence of random or systematic errors in the input parameters ($v_z$, $\ve{B}$, and $\dot{\ve{B}}$) given by the Stokes inversion, the effects of having a finite time sampling when determining $\dot{\ve{B}}$ observationally, and the existence of a finite magnetic diffusion. All of these additional sources of uncertainty will likely overshadow the 1\% uncertainty of the idealized scenarios investigated here.

\begin{acknowledgements}
The work presented in this paper has been funded by a grant from the Deutsche Forschung Gemeinschaft (DFG): project 538773352. This research has made use of NASA's Astrophysics Data System. The authors are grateful to Petri K\"apyl\"a, Maria Kazachenko, Brian Welsch, and Oskar Steiner for useful discussions on the topic. 
\end{acknowledgements}

\bibliographystyle{aa}
\bibliography{ms}

\clearpage
\begin{appendix}
\section{Two-dimensional CO5BOLD simulations}
\label{sec:cobold}

We used the CO5BOLD code \citep{freytag2012simulations} in the box-in-a-star setup to create a realistic 2D atmosphere. The code solves the equations of ideal MHD for a fully compressible gas in a Cartesian box using a finite-volume formulation with an approximate Riemann type solver \citep[HLLMHD;][]{harten1983upstream, 2005ESASP.596E..65S, 2013MSAIS..24..100S}. The constrained transport method used in CO5BOLD ensures the magnetic field divergence free condition ($\nabla \cdot \mathbf{B}=0$) to machine precision. Although, CO5BOLD solves an ideal MHD induction equation, the effective magnetic diffusivity due to the discretization is found to be around $\eta_{\rm eff} \approx 3.0 \times 10^{10}$ cm$^{2}$ s$^{-1}$ for a computational cell width of 10~km and at the bottom of the photosphere \citep{2022A&A...660A.115R}.
For our model, we used a realistic solar-like equation of state taken from the CIFIST project \citep{2009MmSAI..80..711L}. The radiative transfer proceeds via a short characteristics scheme as described in \citet{2017MmSAI..88...22S}, with gray opacities from the MARCS model atmosphere package \citep{2008A&A...486..951G}, provided in tabulated form as a function of gas pressure and temperature.

For this work, we selected a snapshot from a 3D magnetoconvection simulation computed using CO5BOLD, which was originally obtained by embedding a homogeneous, vertical field of 50 G magnetic flux density in a relaxed model of convection \citep[model \texttt{d3gt57g44v50fc} from][]{2018PhDT.......210C}. We extracted a 2D slice in the $(y,z)$ plane from this snapshot at a location in the $x$-direction that contains a few kilogauss flux concentrations. The selected computational domain spans 9.6 Mm x 2.8 Mm in the $(y,z)$ plane, with a cell size of 10 km in both spatial directions, of which a smaller domain is selected for our study (see Fig.~\ref{fig:B_sim_2D}). In CO5BOLD, all the quantities are defined at cell centers, except for the magnetic field components, which are defined at the centers of their corresponding faces. A constant external gravity field with g = 275 m s$^{-2}$ acts along the $z$-direction. The top boundary is open for fluid flow and outward radiation, with the density decreasing exponentially in the boundary cells outside the domain. The vertical component of the magnetic field ($B_z$) is constant across the top boundary, while the transverse component ($B_y$) drops to zero at the same location.

The bottom boundary conditions for the magnetic fields are the same as for the top boundary. For the side boundaries, we used periodic boundary conditions, except for the magnetic field in the $x$ coordinate direction. For this direction, the $x$ component of the magnetic field is set to zero, and a constant extrapolation applies to the other two components. We note that even though initially $B_x$ and $v_x$ are different from zero, the simulation is advanced for an additional 1000 seconds of solar time, after which $B_x$ and  $v_x$ become negligible.

The bottom boundary for the thermodynamic quantities is set up in such a way that the in-flowing material carries a constant specific entropy of 1.775 × 10$^9$ erg g$^{-1}$ K$^{-1}$ resulting in a radiative flux corresponding to an effective temperature (T$_{\rm eff}$) of $\sim5770$ K.

\end{appendix}

\end{document}